\documentclass[12pt]{article}
\usepackage{graphicx}
\usepackage{epsfig}

\usepackage{amssymb}

\begin{document}

\title{ On the scaling ranges of detrended fluctuation analysis for long-memory correlated short series of data}
\author{Dariusz Grech$^{(1)}$\footnote{dgrech@ift.uni.wroc.pl} and Zygmunt Mazur$^{(2)}$ }
\date{}

\maketitle
\begin{center}
%\begin{flushleft}
(1) Institute of Theoretical Physics, University of Wroc{\l}aw, Pl. M.Borna 9, PL-50-204 Wroc{\l}aw, Poland

(2) Institute of Experimental Physics, University of Wroc{\l}aw, Pl. M.Borna 9, \\PL-50-204 Wroc{\l}aw, Poland
%\end{flushleft}
\end{center}

\hspace{1.5 cm}

\begin{abstract}
We examine the scaling regime for the detrended fluctuation analysis (DFA) - the most popular method used to detect the presence of long memory in data and the fractal structure of time series.
First, the scaling range for DFA is studied for uncorrelated data as a function of length $L$ of time series and regression line coefficient $R^2$ at various confidence levels. Next, an analysis of artificial short series with long memory is performed. In both cases the scaling range $\lambda$ is found to change linearly -- both with $L$ and $R^2$. We show how this dependence can be generalized to a simple unified model describing the relation $\lambda=\lambda(L, R^2, H)$ where $H$ ($1/2\leq H \leq 1$) stands for the Hurst exponent of long range autocorrelated data. Our findings should be useful in all applications of DFA technique, particularly for instantaneous (local) DFA where enormous number of short time series has to be examined at once, without possibility for preliminary check of the scaling range of each series separately.
\end{abstract}
$$
$$


\section{Introduction and description of the method.}

 Detrended fluctuation analysis
 (DFA) \cite{DFA_1,DFA_2,DFA_3} is now considered the main tool in searching for fractal \cite{fractals_1, fractals_2, fractals_3}, multifractal \cite{multifr_1, multifr_2} and long memory effects in ordered data. There is more than one thousand articles published on DFA and its applications so far.
The detrended technique has been widely applied to various topics, just to mention: genetics (see e.g. \cite{DFA_2,gen_1,gen_2,gen_3}, meteorology (see e.g. \cite{meteo_1,meteo_2,meteo_3}), cardiac dynamics (see e.g. \cite{heart_1,heart_2}), astrophysics (see e.g. \cite{astro}), finances (see e.g. \cite{finance_1,finance_2,finance_3,finance_4,finance_5,finance_6,finance_7}) and many others.
  The indisputable advantage of DFA over other available methods searching for the Hurst exponent $H$ \cite{Hurst1,Hurst2} in series of data, like the rescaled range method (R/S) \cite{multifr_1, Hurst1, Hurst2, R/S}, is that DFA is shown to be resistant to some extent to non-stationarities in time series \cite{DFA_nonstat}.

 We will not describe the DFA technique in details here, for it is done in many other publications (see e.g. \cite{DFAdescr1,DFAdescr2,DFAdescr3,DFAdescr4}). Instead, we will focus mainly on the issues which are relevant for the so called scaling range being the goal of this article.

Briefly, the DFA method contains the following steps: (i) the time series $x(t)$ ($t=1,2,...,L$) of data  (random walk) is divided into non-overlapping boxes (time windows) of length $\tau$ each, (ii)  the linear trend\footnote{the subtracted trend can also be mimicked by nonlinear polynomial function of order $k$ in so called DFA-$k$ schemes - we will not discus this issue in details here}
%, nevertheless some concluding remarks in the end of this paper refer also to non-linear case}
is found within each box and then subtracted from the signal giving so called detrended signal $\hat{x}(t)$, (iii)  the mean-square fluctuation $F^2(\tau)$ of the detrended signal is calculated in each box and then $F^2(\tau)$ is averaged over all boxes of size $\tau$, (iv) the procedure is repeated for all box sizes $\tau$ ($1<\tau<L$).

 One expects that the power law

 \begin{equation}
 \langle F^2(\tau)\rangle_{box}\sim \tau^{2H}
 \end{equation}
is fulfilled for stationary signal\footnote{this property holds also for non-stationary, positively autocorrelated ($H>1/2$) time series \cite{DFA_nonstat}} where $\langle.\rangle_{box}$ is the expectation value - here, the average taken over all boxes of size $\tau$. The  latter equation allows to make the linear fit in log-log scale to extract the value of $H$ exponent necessary in various applications.
One can also  look alternatively at the above relationship  as a link between the variance of the detrended random walk $\hat{x}(t)$ and its duration time $t$, i.e. $\langle \hat{x}^2(t)\rangle \sim t^{2H}$ what reflects the precise definition of Hurst exponent in stochastic processes. The $H$ exponent clearly indicates the randomness nature of this process.
One deals with uncorrelated steps in data series if  $H=1/2$, once for other values of $H$ these steps are respectively anticorrelated ($0<H<1/2$) or autocorrelated with (positive) long memory ($1/2\leq H \leq 1$).

The edge part of time series is usually not covered by any box. Some authors suggest to overcome this difficulty performing DFA in two opposite directions in time series, i.e. according to increasing and then according to decreasing time arrow (see e.g. \cite{kantelhardt}). The average of mean-square fluctuations from such divisions is then taken for evaluation of time series properties.

 We proposed another solution in Refs.\cite{grech_1,grech_2}. If the remaining part of time series $\Delta L$ has the length $\tau/2\leq\Delta L<\tau$, we cover it by an additional box of size $\tau$ partly overlapping the preceding data. If $\Delta L<\tau/2$, we do not take into account the part of data contained in $\Delta L$. Such recipe is particularly useful in the 'local' version of DFA \cite{finance_1,grech_1,grech_2,DFA_loc2, DFA_loc4,grech_3,kristoufek}, where the time arrow is important. Throughout this article we will apply the latter approach.

 If time series are infinitely long, the formula in Eq.(1) holds for all $\tau's$. However, in practise we always deal with finite, and sometimes with rather short time series. Particularly, it is a case for the mentioned already instantaneous or local DFA analysis, where one wants to find a dynamics of fractal properties changing in time and (or) their time dependent long memory in data. Covering the data series with boxes, we are finally stuck with situation that for small number of boxes covering the time series (for large $\tau$ ), the scaling  is not revealed in Eq.(1) due to small statistics we deal with. In other words, we are allowed in this case to take $\tau$ only within some range $\tau_{min}\leq \tau\leq \tau_{max}$ called the scaling range. One expects within this range "sufficiently good" performance of the power law, thus leading to $H$ exponent extraction via linear fit. But what does this "sufficiently good" performance exactly mean? In most research activities authors end up with $\tau_{max}\sim 1/4 L$, where $L$ is the total length of considered data. Is it still good or already too large scaling range? This problem is somehow circumvented in papers but it does have impact on the final results. The aim of this and other forthcoming article \cite{grech_fut} is to confront this issue. Our approach will be different than the one published in \cite{stanley_scr,michalski_scr,comp_meth_fluct_anal}. The goal is to find qualitative and quantitative dependence between the scaling range $\lambda \equiv \tau_{max}$ and main parameters of time series like its length, level of long memory described by the Hurst exponent $H$, and the goodness of linear fit induced by the form of Eq.(1) in log-log scale. The latter one is usually measured by the $R^2$ regression line coefficient. All this can be done at desired confidence level ($CL$) indicating the minimal ratio of time series fulfilling the functional dependence $\lambda = \lambda (L, R^2, H)$. We are going to find this relation below.

 Throughout this paper we assumed that $\tau_{min} = 8$ because below this threshold a significant lack of scaling in DFA is observed due to emergence of artificial autocorrelations associated with too short bursts of data in $\tau$ boxes. We start with analysis of uncorrelated data in the next section and proceed with long memory correlated time series in section 3. Section 4 tries to obtain an unified formula for scaling range vs $L$ and $R^2$ for all $H\geq 1/2$. Although the presented considerations are done exclusively for DFA method, they can be easy extended to other detrended methods introduced in literature, in particular to those based on moving averages \cite{DMA1,DMA2,DMA3,DMA4}. The latter analysis is left to another publication [40].

  %It is a topic of forthcoming publication \cite{grech_fut}.

\section{DFA scaling ranges for uncorrelated data}

The starting point for the entire search is the statistical analysis of an ensemble of artificially generated time series with a given length. For this ensemble we find the percentage rate of series which  are below the specified level of regression line fit parameter $R^2$. This rate will obviously depend on the maximum size of the box $\tau_{max}$. The larger $\tau_{max}$, the percentage rate of series not matching the assumed criterion for $R^2$ will be also larger. Fig.1. illustrates this fact for two specified series lengths $L=10^3$ and $L=3\times 10^3$ of uncorrelated data increments ($H=1/2$) drawn from the normalized Gaussian distribution. The rejection rate, i.e the percentage rate of series not matching the assumed criterion for $R^2$ is shown there for different $R^2$ values.

We took two particular values of rejection rate in further analysis: $2.5\%$ and $5.0\%$, connected with confidence levels $CL= 97.5\%$ and $CL=95.0\%$ respectively.
All data have been gathered numerically on a set of $5\times 10^4$ artificially generated time series of length between $5\times 10^2\leq L\leq2\times 10^4$ for the above-stated confidence levels. The $\tau_{max}$ value corresponding to required $CL$ and for given $R^2$ is identified with the scaling range and referred to exactly as $\lambda$.

Introducing for convenience a new parameter $u=1-R^2$, we may search for a $\lambda(L)$  dependence for $L\leq 2\times 10^4$, for different values of $u$ and for selected $CL$'s. The results are presented in a series of graphs in Figs. 2, 3 and reveal a very good linear relationship between the scaling range profile and the length of uncorrelated data\footnote{obviously $\lambda(L,u) \in \textsc{Z}$, so in fact only the integer part of RHS of Eq.(2) should be taken for determination of $\lambda(L,u)$}:

\begin{equation}
\lambda(u,L) = A(u)L + B(u)
\end{equation}
The functional dependence of coefficients $A(u)$ and $B(u)$  on $u$ has to be further specified from the regression line fit of the above equation. The latter procedure yields to the values of $A$ and $B$ estimated for the spread of $u$ parameters %($u=n\times 10^{-2},  n=1,2,...,5$)
and gathered in Fig.4.

We see from these graphs that the dependence of $A(u)$ is again linear for both cases of $CL=97.5\%$ and $CL=95\%$, while the value of B varies very weakly with $u$, what legitimates us to accept $B(u) = b = const$.

Ultimately, the foregoing considerations lead to the following simple formula describing the full scaling range dependence on $L$ and $u$:
\begin{equation}
\lambda(u,L) = (au+ a_0)L + b
\end{equation}
with some unknown constants $a$, $a_0$ and $b$ to be fitted.

We made the fit for Eq.(3) requiring minimization of mean absolute error (MAE) and simultaneously, minimization of the maximal relative error (ME) for each of the fitting points\footnote{{we considered $u_j=5\times 10^{-3}(1+j)$ where $j=1,2,...,9$ and $L_i$ covering the range from $L=5\times10^2$ up to $L=2\times 10^4$ as indicated on plots}}$(L_i, u_j)$. The MAE denoted as $\Delta_{MAE}(\lambda)$ is understood as
\begin{equation}
\Delta_{MAE}(\lambda)=1/N_{(ij)} \sum_{ij}|(\lambda^{exp}_{ij}(L,u)-\lambda_{ij}(L,u))/\lambda_{ij}(L,u)|
\end{equation}
where $\lambda_{ij}(L,u)\equiv \lambda(L_i,u_j)$ is taken from Eq.(3) for the particular choice $L=L_i$ and $u=u_j$, while $\lambda^{exp}_{ij}(L,u)$ is the respective value simulated numerically for given ensemble of time series, and $N_{(ij)}$ counts different $(ij)$ pairs.

Similarly ME marked below as $\Delta_{ME}$ is simply defined as
\begin{equation}
\Delta_{ME} = \max_{(ij)}(|\lambda^{exp}_{ij}(L,u)-\lambda_{ij}(L,u))|/\lambda_{ij}(L,u))
\end{equation}
Note that some pairs $(L_i,u_j)$ are not permitted by the specific $CL$ demand\footnote{we found that following pairs: $(L<3000, u_1), (L<1800, u_2), (L<1500, u_3), (L<1000, u_4)$,\\ $(L<1000, u_5)$, $(L<800, u_6)$, $(L<800, u_7)$, $(L<600, u_8)$ do not match the $CL=97.5\%$ requirement, and: $(L<2400, u_1), (L<1800, u_2), (L<1200, u_3), (L<1000, u_4), (L<800, u_5), (L<800, u_6), (L<600, u_7)$ do not match the $CL=95\%$ demand}. It is seen already in Fig.1. These points are therefore absent in Figs 2-3,\,\,5-6.

The fitting procedure led to the values of parameters in equation Eq.(3) gathered in Table 1. The exemplary results of scaling ranges for the wide spread of $L$ and $u$ values are presented graphically in Figs.\,5,\,6.
Whenever $\lambda(L,u)$ comes out negative in the found fitting patterns for the particular length of the series, one should interpret this as a lack of scaling range at the given confidence level $CL$ for the required value of regression line coefficient $R^2$ within DFA.

\section{DFA scaling range for long-memory correlated data}

The analysis presented in the previous section can be extended to time series manifesting  long memory. The series with $0.5<H<0.9$ are of particular interest since they correspond to long-range autocorrelated data one often meets in practice in various areas.

To construct such signals we used Fourier filtering method  (FFM) \cite{ffm} . The level of autocorrelations in this approach was directly modulated by the choice of autocorrelation function $C(\delta t)$ which satisfies for stationary series with long memory the known power law \cite{corr-gamma}:
\begin{equation}
C(s)\equiv \langle \Delta x(t+s)\Delta x(t) \rangle \sim H(2H-1)s^{2H-2}
\label{corr}
\end{equation}
where $\Delta x(t)=x(t+1)-x(t)$, ($t=1, 2,..., L-1$) are increments of discrete time series, $s$ is the time-lag between observations, $H$ is the Hurst exponent \cite{Hurst1,Hurst2}, and the average $\langle\rangle$  is taken over all data in series.

We start with similar analysis as the one shown in Fig.1 for uncorrelated data. Fig.7 presents an example of plot made for the ensemble of $5\times 10^4$  autocorrelated signals of length $L=10^3$ with $H=0.7$. The percentage rate of rejected time series not satisfying the assumed goodness $R^2$ of DFA fit is shown there for several distinct $R^2$ as a function of maximal box size $\tau_{max}$. The outcome of such analysis for a range of simulated data lengths and for various Hurst exponents can be collected in number of plots as in Figs.8a,\,9a for $\lambda(L)$,  and in Fig.8b,\,9b for $\lambda(u)$ dependence. To make the figure readable and due to lack of space, only plots for $u=0.02$ and $L=10^3$ are shown. The relations for other values look qualitatively the same. We should not be surprised, taking into account the results of the previous chapter, that these relationships are again linear. Thus the formula in Eq.(2) is more general and coefficients $A(u)$ and $B(u)$ are linear function of $u$ also for series with memory.
The latter relationships are drawn in details for $H=0.6,\, 0.7,\, 0.8$ in Fig.10. In particular, we notice from Fig.\,10b the similar behavior of $B(u)$ coefficient for autocorrelated data as it has been observed in the previous section for uncorrelated signals, i.e. $B(u)$ remains almost constant as a function of $u$. Moreover, its dependence on $H$ is also negligible. Thus, the formula postulated in Eq.(3) applies also for autocorrelated data with $a$, $a_0$ and $b$ coefficients to be fitted independently for each $H$.

We did such a fit for series with long memory, assuming the same criterions for MAE and ME as previously. The results are collected in Table 1 for two different confidence levels and are shown graphically in Figs.\,11-16. These figures generalize plots shown for $H=0.5$ in Figs.\,5,\,6. The extremely good linear relationship of Eq.(3) is kept for autocorrelated signals up to $L=10^4$. Only for highly autocorrelated series ($H> 0.8$) or very long ones ($L\geq 10^4$) we noticed some slight departure from the linear dependence\footnote{the predicted scaling ranges from Eq.(3) were nevertheless lower in these cases than the ones coming from the direct simulation}.

\begin{table}
\centering
\begin{tabular}{||c||c|c|c|c|c||c|c|c|c|c||}
   \hline
   $H\setminus CL$ & $a^{97.5\%}$ & $a^{97.5\%}_0$ & $b^{97.5\%}$ & $\Delta^{97.5\%}_{MAE}$ & $\Delta^{97.5\%}_{ME}$ & $a^{95\%}$ & $a^{95\%}_0$ & $b^{95\%}$ & $\Delta^{95\%}_{MAE}$ & $\Delta^{95\%}_{ME}$\\
\hline
$H=0.5$ & 6.02 & 0.0034 & -92  & 1.8\% & 5.1\% & 7.00  & 0.0031  & -100 & 1.8\% & 5.1\%\\
\hline
$H=0.6$ & 6.14 & 0.0110 & -95 & 1.6\% & 5.2\% & 7.22 & 0.0098 & -105 & 1.9\% & 5.3\%\\
\hline
$H=0.7$ & 6.46 & 0.0124 & -97 & 1.9\% & 4.6\% & 7.66 & 0.0084 & -103 & 1.8\% & 4.6\%\\
\hline
$H=0.8$ & 6.88 & 0.0136 & -100 & 1.5\% & 4.8\%  & 8.12 & 0.0091 & -104 & 2.5\% & 6.0\%\\
\hline
\end{tabular}
\caption{Results of the best fit for coefficients in Eq.(3) found for series with various autocorrelation level measured by $H$ exponent and for chosen two confidence levels: $97.5\%$ and $95\%$. The accuracy of fitted parameters are respectivel: $\Delta a=\pm 10^{-3}$, $\Delta a_0 =\pm 10^{-5}$, $\Delta b=0$.}
\label{tab1}
\end{table}

\section{Towards unified model of scaling ranges}

 Finally, we should investigate if there exists a unified formula with the minimal number of free parameters, able to describe all scaling ranges of both uncorrelated and autocorrelated data. So far we know that Eq.(3) with parameters fitted according to Table 1 describes very well $\lambda(L,u)$ dependence for given $H$. We should discuss then the form of relationships $a(H)$, $a_0(H)$, and $b(H)$ in the relation

 \begin{equation}
 \lambda(u,L,H) = (a(H)u+ a_0(H))L + b(H)
 \end{equation}

 Looking at the bottom panels of Figs.4, 10 one perceives immediately  that the assumption  $b(H) = const$ can be justified.
  Similarly, we may easily notice from data collected in Table 1 that $a_0 (H)/({a(H)u}) \lesssim\textit {O}({10^{-1}})$. It means that the component $a(H)u$ gives the leading contribution to the linear factor $a(H)u + a_0(H)$ in Eq. (7) for each value of $H$ and therefore, one should focus mainly on $a(H)$ dependence depicted in Fig.17. The latter relationship also appears to be linear, which allows to represent Eq.(7) in its simplest unified form containing the smallest number of four free parameters ($\alpha, \beta, \alpha_0, \gamma$) as follows:

\begin{equation}
 \lambda(u,L,H) = ((\alpha H + \beta)u+ \alpha_0)L + \gamma
 \end{equation}
Demanding minimization of MAE and ME during fitting procedure of the proposal given in Eq.(8) to all data points $\lambda^{exp}(L_i,u_j,H)$ indicated in previous sections, we arrive with the best fit results for these free parameters as shown in Table 2. The obtained unified formula can be particularly useful while doing interpolation to arbitrary autocorrelation levels $1/2<H<1$.

In fact the fit based on Eq.(8) is of the same quality as the one produced by Eq.(3) (see Table 1 and 2 to compare MAE and ME errors).
The difference between two fitting methods is so negligible that it cannot be noticed graphically. Therefore the fitting lines shown in series of Figs.11-16 describe equally well the unified model based on data from Table 2 and the 'local' fit based on data from Table 1.
We may also easy conclude from Eq.(8) that the average relative change in the scaling range $\delta \lambda(\delta H)/\lambda (H)$ due to the small change $\delta H$ in Hurst exponent is given as

\begin{equation}
 \frac{\delta\lambda(\delta H,u)}{\lambda(H,u)} \simeq \alpha H u
 \end{equation}
 and varies from $3\%$ (at $R^2=0.99$) to $10\%$ (at $R^2=0.97$) for any change $\delta H=0.1$ in the investigated signal.

\begin{table}
\centering
\begin{tabular}{||c|c|c|c|c|c||}
   \hline
    $\alpha^{97.5\%}$ & $\beta^{97.5\%}$ & $\alpha^{97.5\%}_0$ & $\gamma^{97.5\%}$ & $\Delta^{97.5\%}_{MAE}$ & $\Delta^{95\%}_{ME}$\\
    \hline
3.40 & 4.16 & 0.0097 & -96  & 1.9\% & 5.9\% \\
\hline
 $\alpha^{95\%}$ & $\beta^{95\%}$ & $\alpha^{95\%}_0$ & $\gamma^{95\%}$ & $\Delta^{95\%}_{MAE}$ & $\Delta^{95\%}_{ME}$\\
\hline
3.95  & 5.03  & 0.0070 & -106 & 2.7\% & 5.7\%\\
\hline
\end{tabular}
\caption{Results of the best fit for coefficients in unified formula in Eq.(8). Fit was done for all data coming from investigated series, separately for two chosen  confidence levels: $97.5\%$ and $95\%$. The accuracy of fitted parameters have been estimated: $\Delta \alpha = \Delta \beta = 10^{-3}$, $\delta \alpha_0 = 10^{-5}$, $\Delta \gamma =0$.}
\label{tab1}
\end{table}

\section{Discussion and Conclusions}

In this study we searched for the scaling range properties of the most substantial power law between
fluctuations of detrended random walk $F^2(\tau)$ and the length of the time window $\tau$ in which such
fluctuations are measured. This power law proposed within DFA technique gives us an important information
about the nature of randomness in stochastic process via link between the scaling exponent $H$ and the
autocorrelation exponent between steps of random walk. Therefore, the precise knowledge of scaling range
dependence on any other involved parameters  is a substantial task and has an impact on the final outcomes of
DFA power law quoted in Eq.(1). We did our simulations on the ensemble of $5\times 10^4$ short and medium-length time series with $5\times10^2\leq L \leq 10^4$. We varied also their autocorrelation properties in order to  reflect properties of real random walk signals mostly existing in nature.

First, it has been found that for uncorrelated process, the scaling range $\lambda$ of DFA power law is
the perfect linear function of data length and the goodness of linear fit to power law formula in Eq.(1).
Moreover, this linear relationship extends also to time series with long memory. The uniform shape of
 $\lambda (L, u)$ dependence for different memory levels in data, rises the question if one
unified simple formula describing dependence of scaling range on all parameters in a game, i.e. $\lambda
(L, u, H)$ exists . We found such a formula, and showed that
it fits data obtained from numerical simulations no worse than patterns previously found in this article for
$\lambda (L, u)$ dependence at separate values of H. The unified formula contains only four free
parameters, which were calculated with high precision and are presented in Table 2. We showed also that
scaling range grows with a long memory level present in time series -- on the average of $3\div10\%$ for every $\delta H = 0.1$ (see Eq.(9)). A rather slight increase in the scaling range for the series with memory in
comparison with the array of uncorrelated data may entitle us to simplify the scaling range
for the series with long memory, using a model for uncorrelated data, i.e. with $H = 1/2$.
%We have also checked that the scaling range slightly increases in the case of nonlinear (polynomial) trends subtracted from time series data within so called DFA-$k$ technique.
The presented results can be considered therefore as the lower limit for the DFA scaling range profile.

The relations we found strike with their simplicity and make a useful recipe how to determine the scaling ranges, especially for short time series -- wherever one needs to consider very large data sets
arranged in shorter subseries. In particular, these results can be used in search for evolving (time-dependent) local Hurst exponent in large amount of moving time windows. The extension of this approach to other techniques of fluctuation analysis (FA) can also be done [40].

\newpage

\begin{figure}
\begin{center}
{\psfig{file=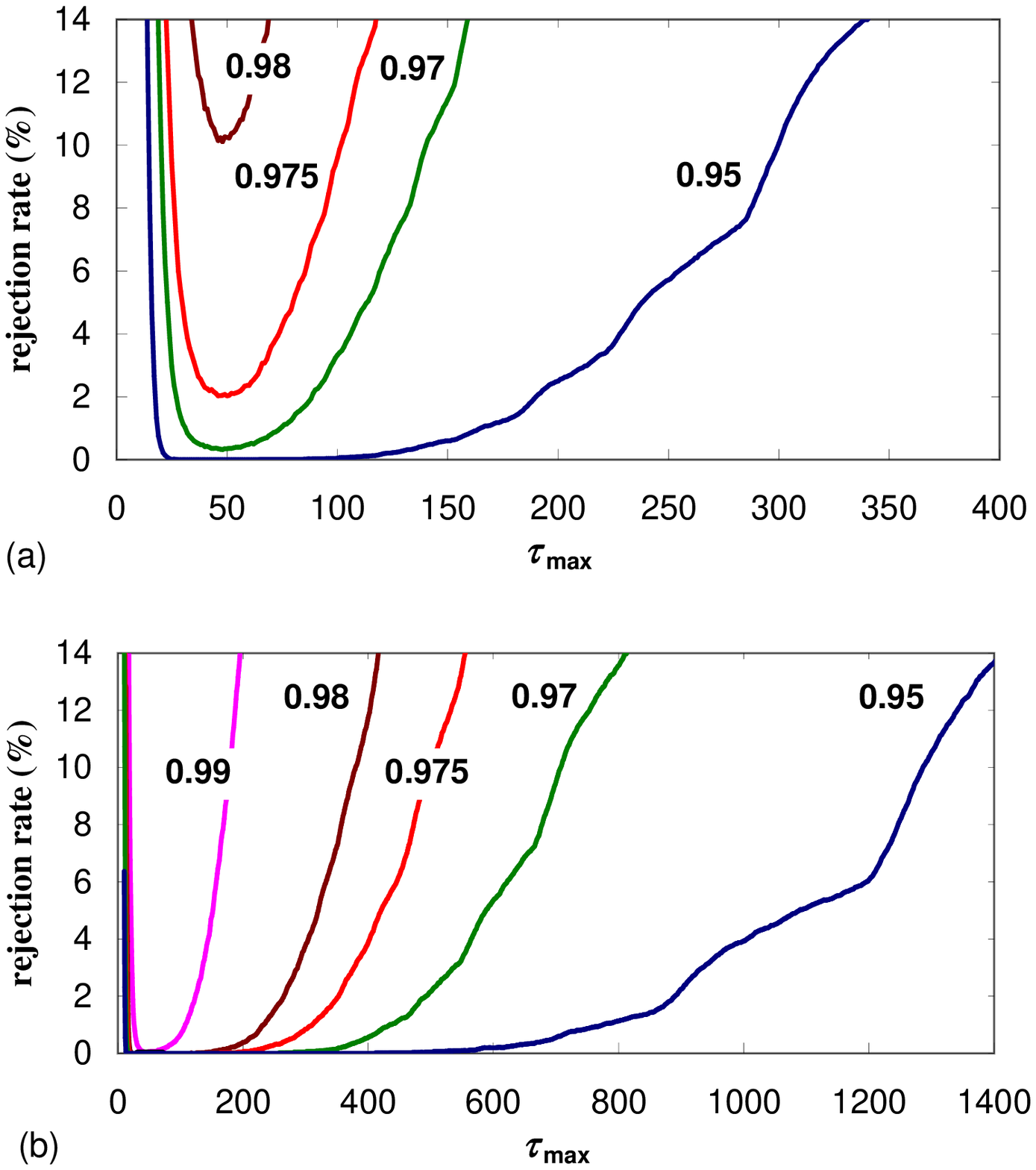,width=15cm}}
\end{center}
\caption{Percentage rate (\%) of rejected time series with scaling range not providing indicated goodness of regression line fit $R^2$ in DFA procedure. Results are based on the ensemble of $5\times 10^4$ time series and are drawn as a function of maximal box size $\tau_{max}$ for two lengths of uncorrelated data: (a) $L=10^3$ and (b) $L=3\times 10^3$.}
\end{figure}

\begin{figure}
\begin{center}
{\psfig{file=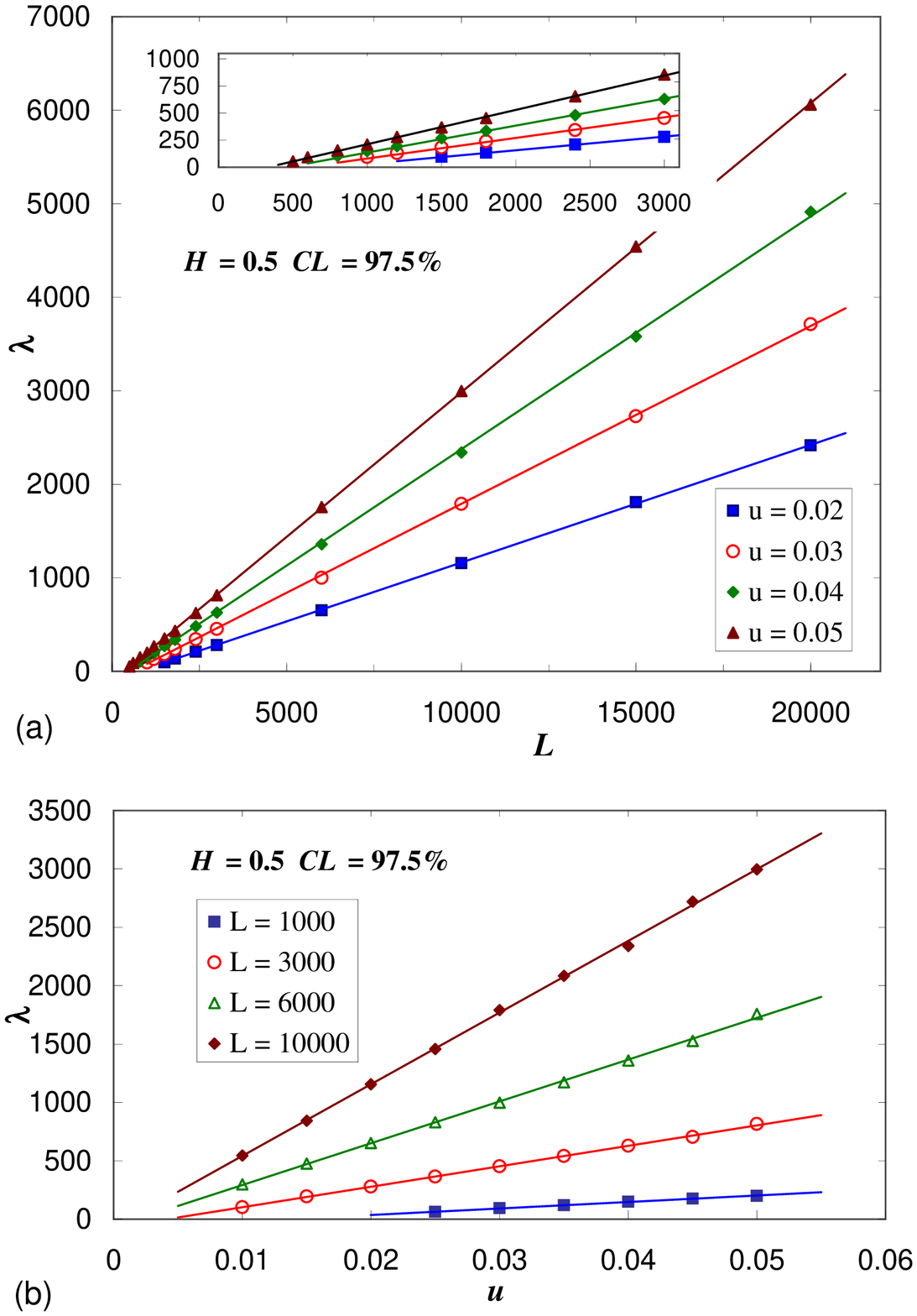,width=14cm}}
\end{center}
\caption{Dependence between scaling range $\lambda$ as in Fig.1 and: (a) time series length $L$  or (b) the goodness of linear fit $u=1-R^2$ .
 Examples of $\lambda(L, u=fixed)$ and $\lambda(u, L=fixed)$ relations for various $u$ and $L$ values are shown at $97.5\%$ confidence level.}
\end{figure}

\begin{figure}
\begin{center}
{\psfig{file=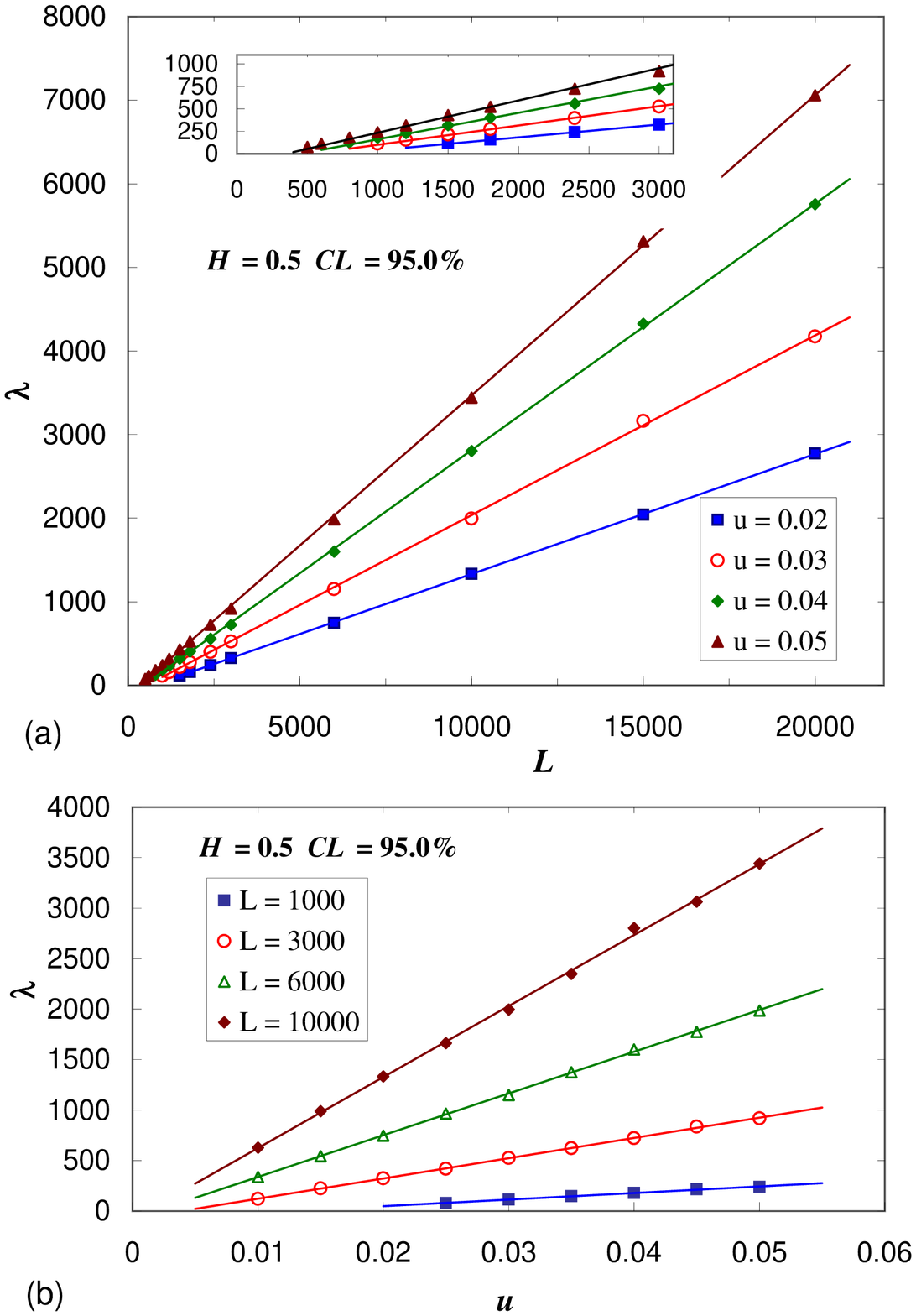,width=14.5cm}}
\end{center}
\caption{Same as in Fig.2 but at $95\%$ confidence level.}
\end{figure}

\begin{figure}
\begin{center}
{\psfig{file=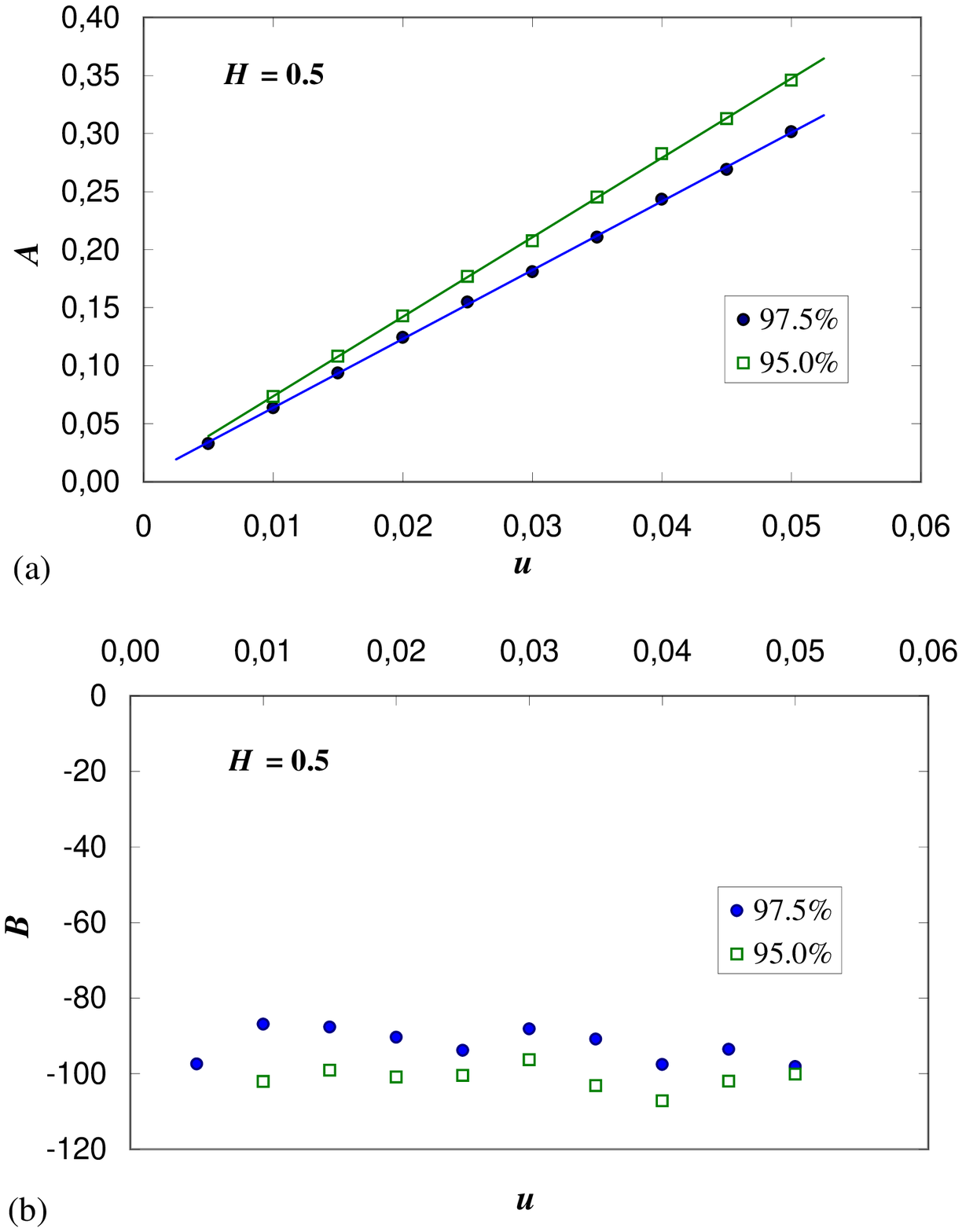,width=15cm,angle=0}}
\end{center}
\caption{(a) Dependence between fitted coefficient $A$ in Eq.(2) and the goodness of DFA fit $u=1-R^2$ for uncorrelated data ($H=0.5$) shown for two distinct confidence levels: $CL=97.5\%$, $CL=95.0\%$. (b) Same for $B$ coefficient in Eq.(2).}
\end{figure}

\begin{figure}
\begin{center}
{\psfig{file=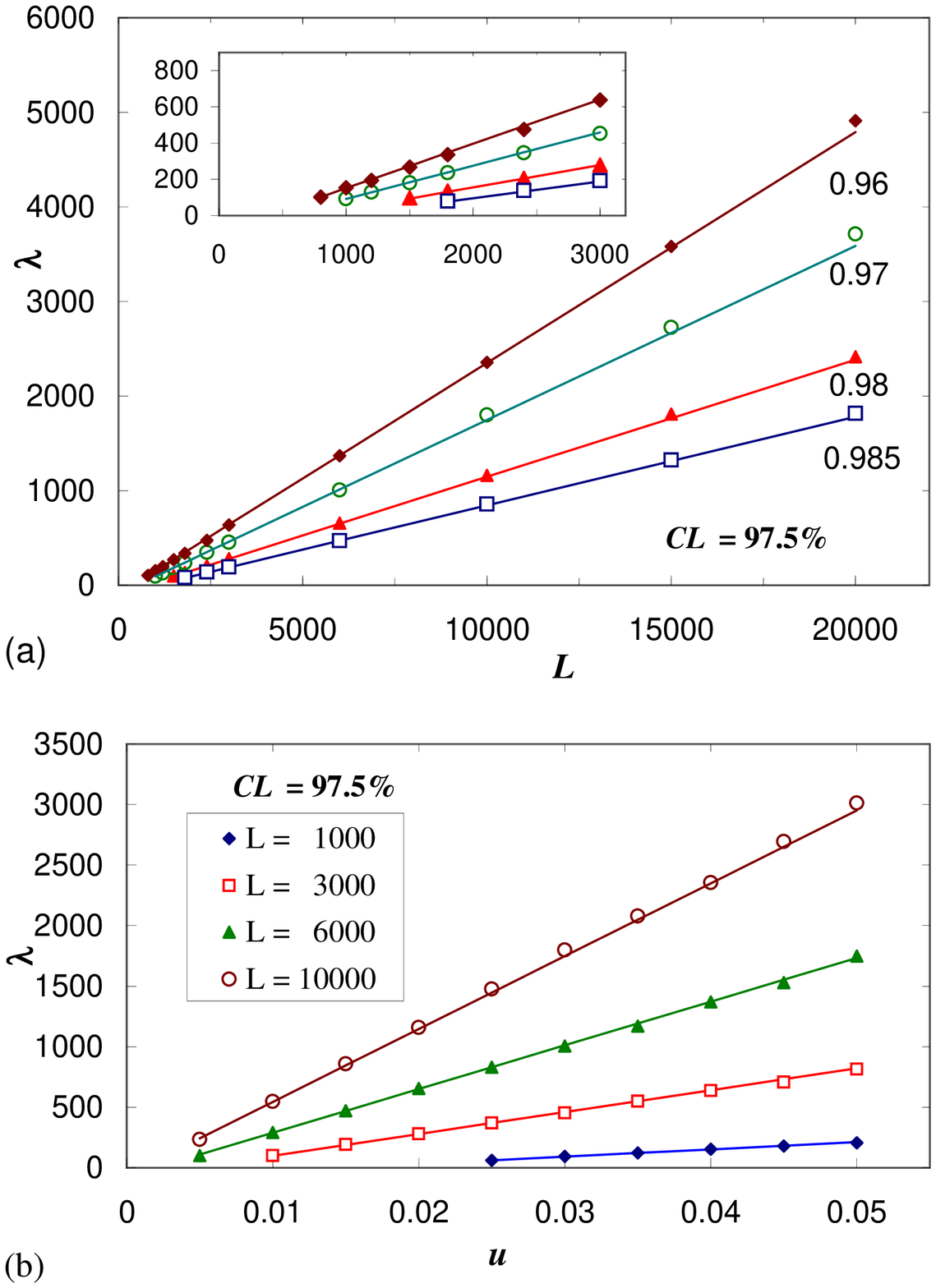,width=14cm,angle=0}}
\end{center}
\caption{Best fit results of the relationship suggested in Eq.(3) at $CL=97.5\%$ level. Continuous lines represent the fit of Eq.(3) to data (marked as points). (a) Results shown for $\lambda(L, u=fixed)$  at $R^2=1-u=0.96,\,0.97,\,0.98,\,0.985$. (b) results for $\lambda(L=fixed, u)$  shown for chosen lengths of uncorrelated data $L=10^3,\,3\times 10^3,\,6\times 10^3,\,10^4$. Parameters of fits are gathered in Table 1.}
\end{figure}

\begin{figure}
\begin{center}
{\psfig{file=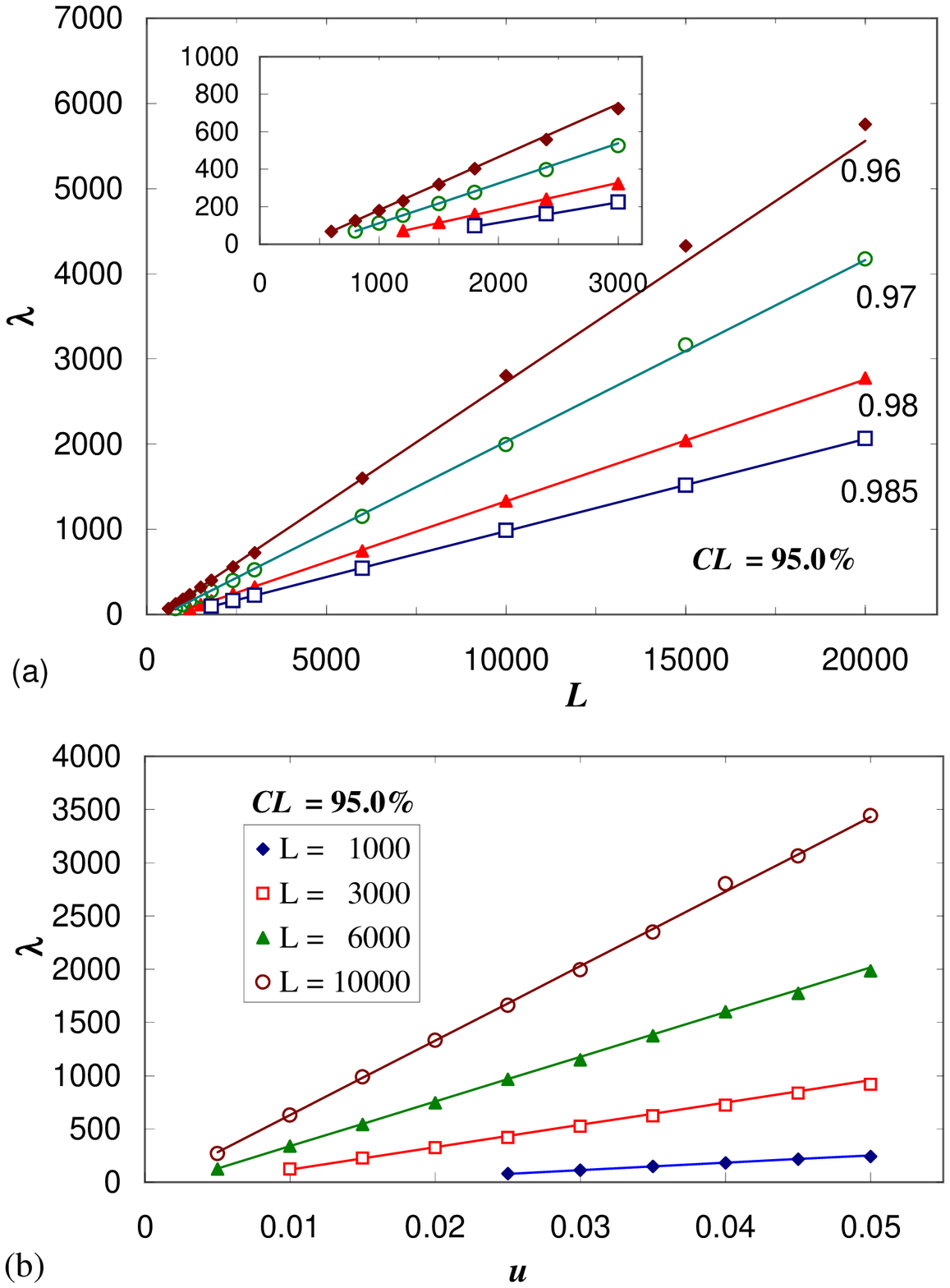,width=15cm,angle=0}}
\end{center}
\caption{Same as in Fig.5 but at $CL=95\%$ level.}
\end{figure}

\begin{figure}
\begin{center}
{\psfig{file=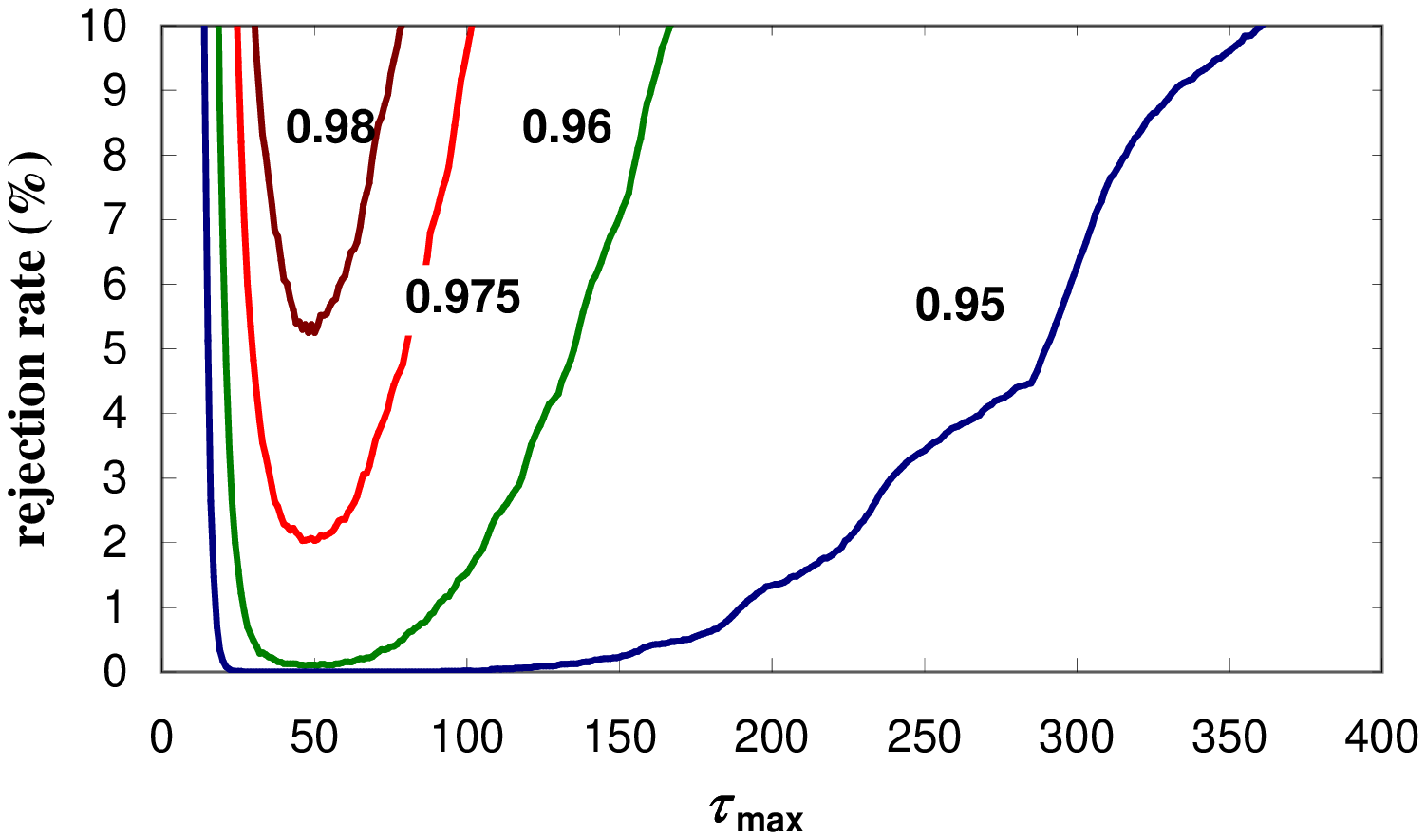,width=15cm,angle=0}}
\end{center}
\caption{Percentage rate (\%) of rejected time series as a function of maximal box size $\tau_{max}$. Exemplary results are based on the ensemble of $5\times 10^4$ time series for autocorrelated signal of length $L=10^3$ with $H=0.7$.}

\end{figure}

\begin{figure}
\begin{center}
{\psfig{file=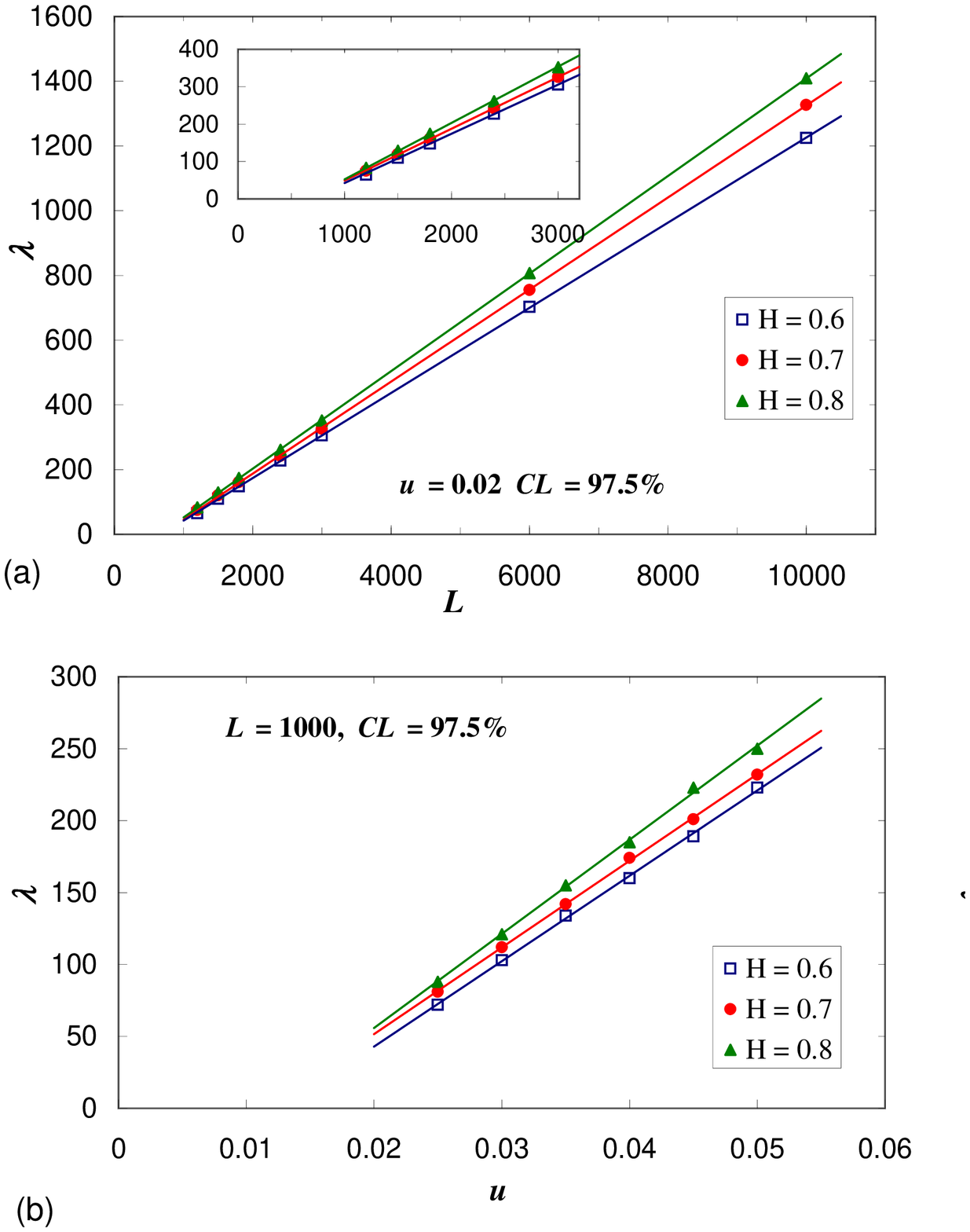,width=14cm,angle=0}}
\end{center}
\caption{Dependence between scaling range $\lambda$ and: (a) time series length $L$  or (b) the goodness of linear fit $u=1-R^2$ for autocorrelated signals. The autocorrelation level is indicated by Hurst exponent $H$. Plots are shown only for $L=10^3$ and $u=0.02$ $(R^2=0.98)$ but the relations $\lambda(L,u)$ for the remaining values of $L$ and $u$ look qualitatively very similar (not shown). The linear relationship $\lambda (L,u)$ is seen at $CL=97.5\%$ level.}
\end{figure}

\begin{figure}
\begin{center}
{\psfig{file=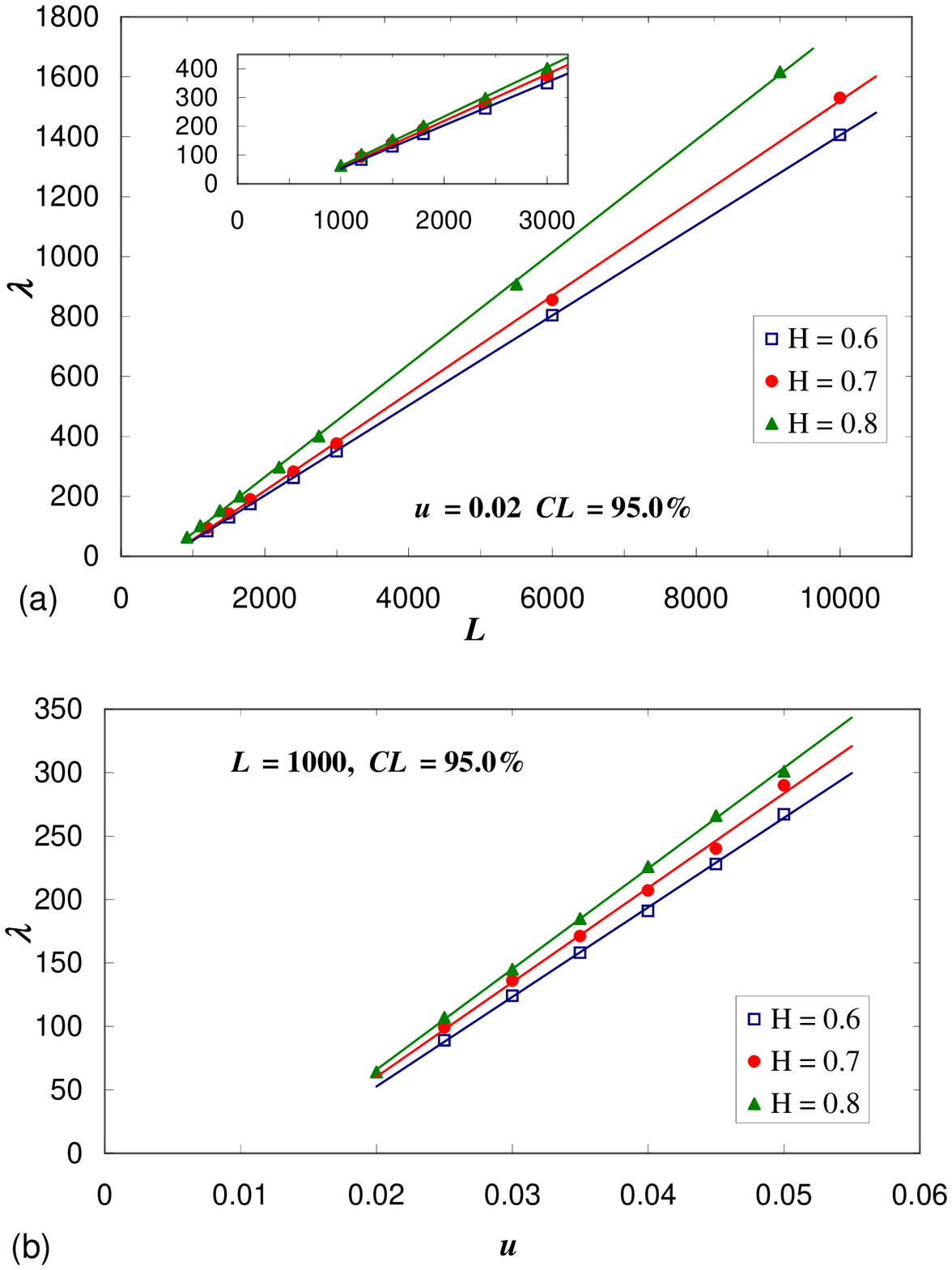,width=15cm,angle=0}}
\end{center}
\caption{Same as in Fig.8 but at $CL=95\%$}
\end{figure}

\begin{figure}
\begin{center}
{\psfig{file=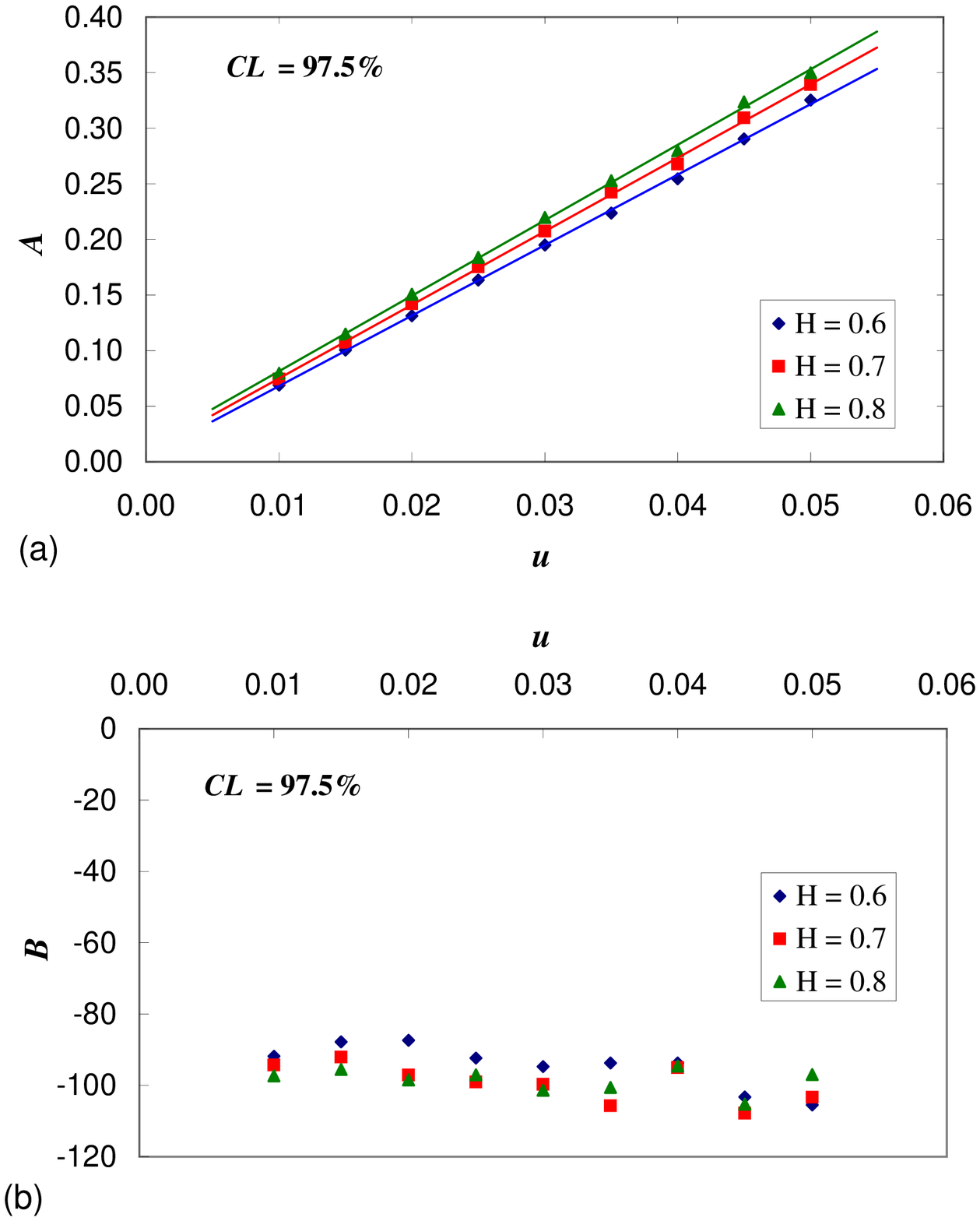,width=15cm,angle=0}}
\end{center}
\caption{(a) Dependence between coefficient $A$ in Eq.(2) and the goodness of DFA fit $u=1-R^2$ for autocorrelated data ($H=0.6, H=0.7, H=0.8$). Confidence level $CL=97.5\%$ is considered but $CL=95\%$ looks similarly (not shown). (b) Same for $B$ coefficient in Eq.(2).}
\end{figure}

\begin{figure}
\begin{center}
{\psfig{file=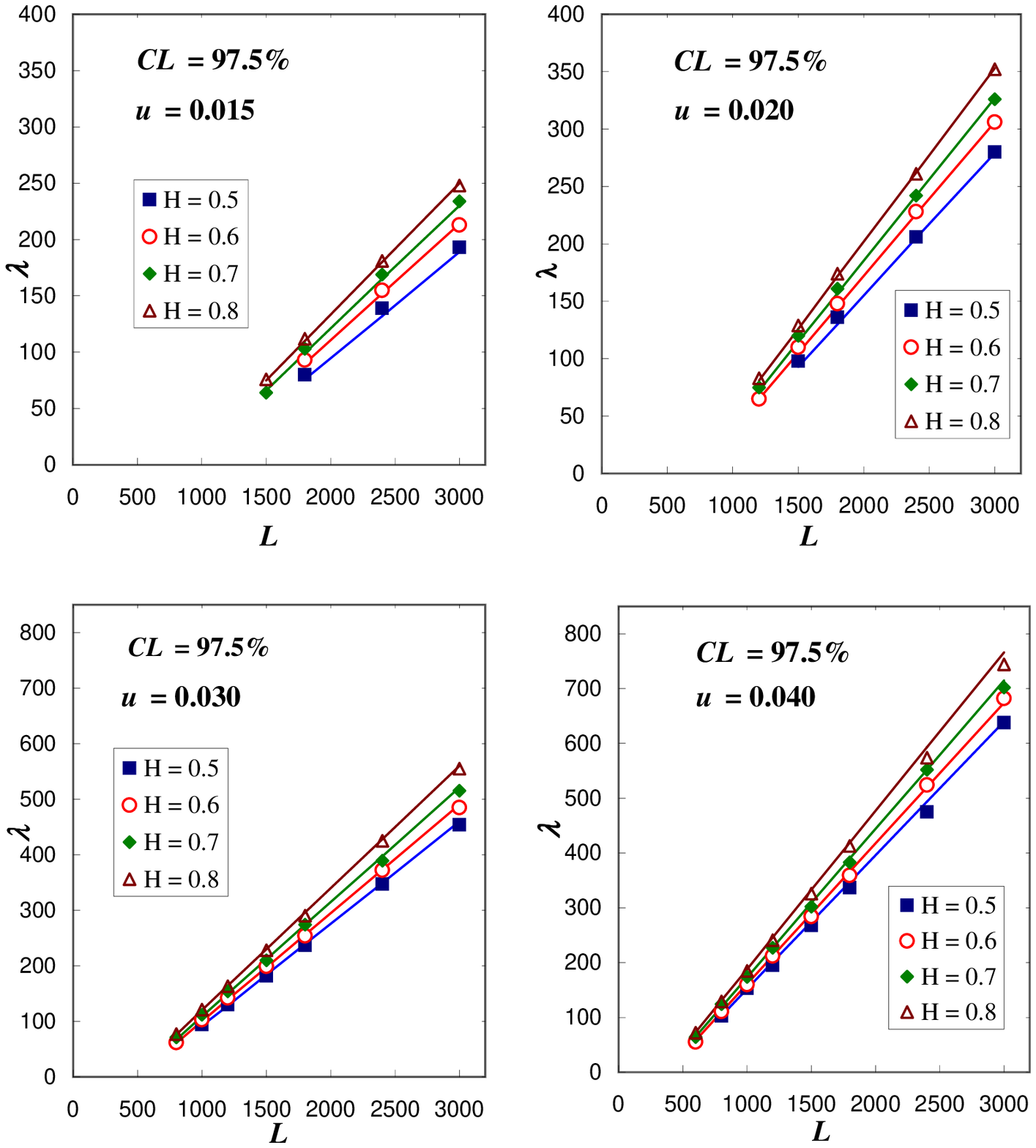,width=15cm,angle=0}}
\end{center}
\caption{Best fit results of Eq.(7) found for simulated series of autocorrelated data at $CL=97.5\%$ level and shown for $L\leq 3000$. Continuous lines represent fit of Eq.(7) to data points marked as dots for chosen $u$. The cases of other $u$ values (not shown due to lack of space) look identically. Parameters of the fit are gathered in Table 2.}
\end{figure}

\begin{figure}
\begin{center}
{\psfig{file=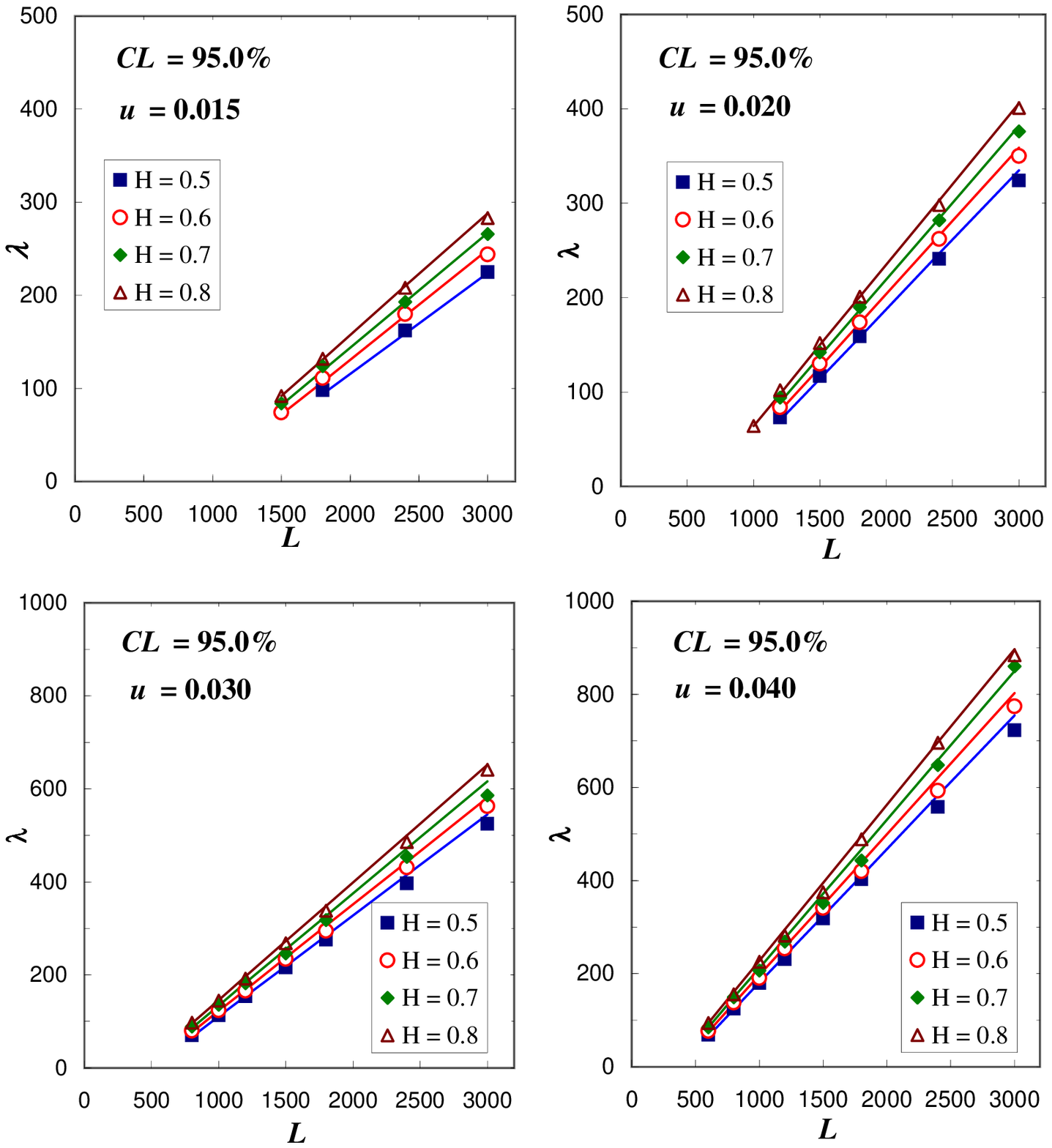,width=15cm,angle=0}}
\end{center}
\caption{Same as in Fig.11 for $CL=0.95\%$.}
\end{figure}

\begin{figure}
\begin{center}
{\psfig{file=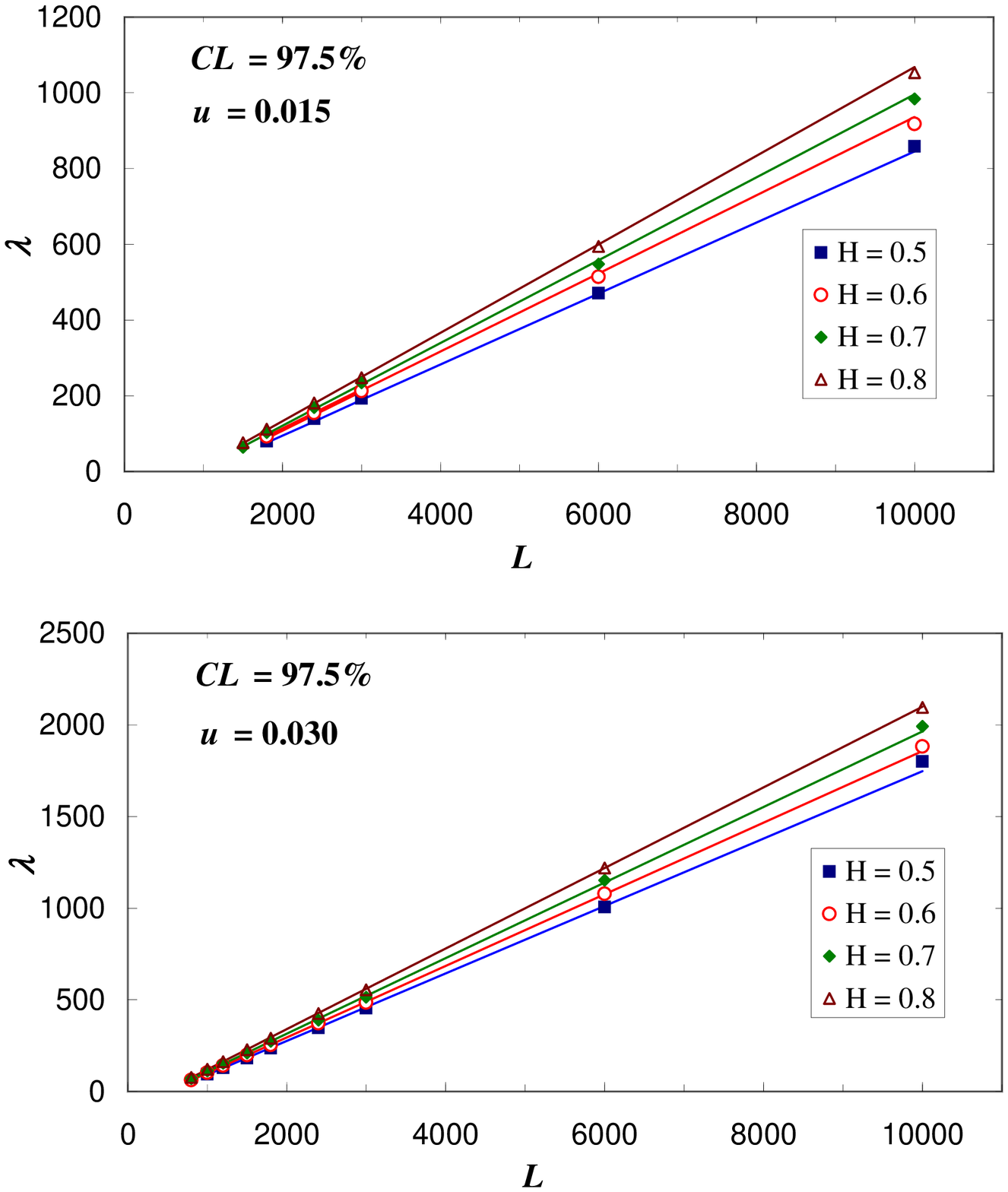,width=15cm,angle=0}}
\end{center}
\caption{Same dependence as in Fig.11 shown for longer series of data ($L\leq 10^4$). Only cases for $u=0.015$ and $u=0.030$ are presented due to lack of space. Plots for other $R^2$ coefficients look qualitatively the same.}
\end{figure}

\begin{figure}
\begin{center}
{\psfig{file=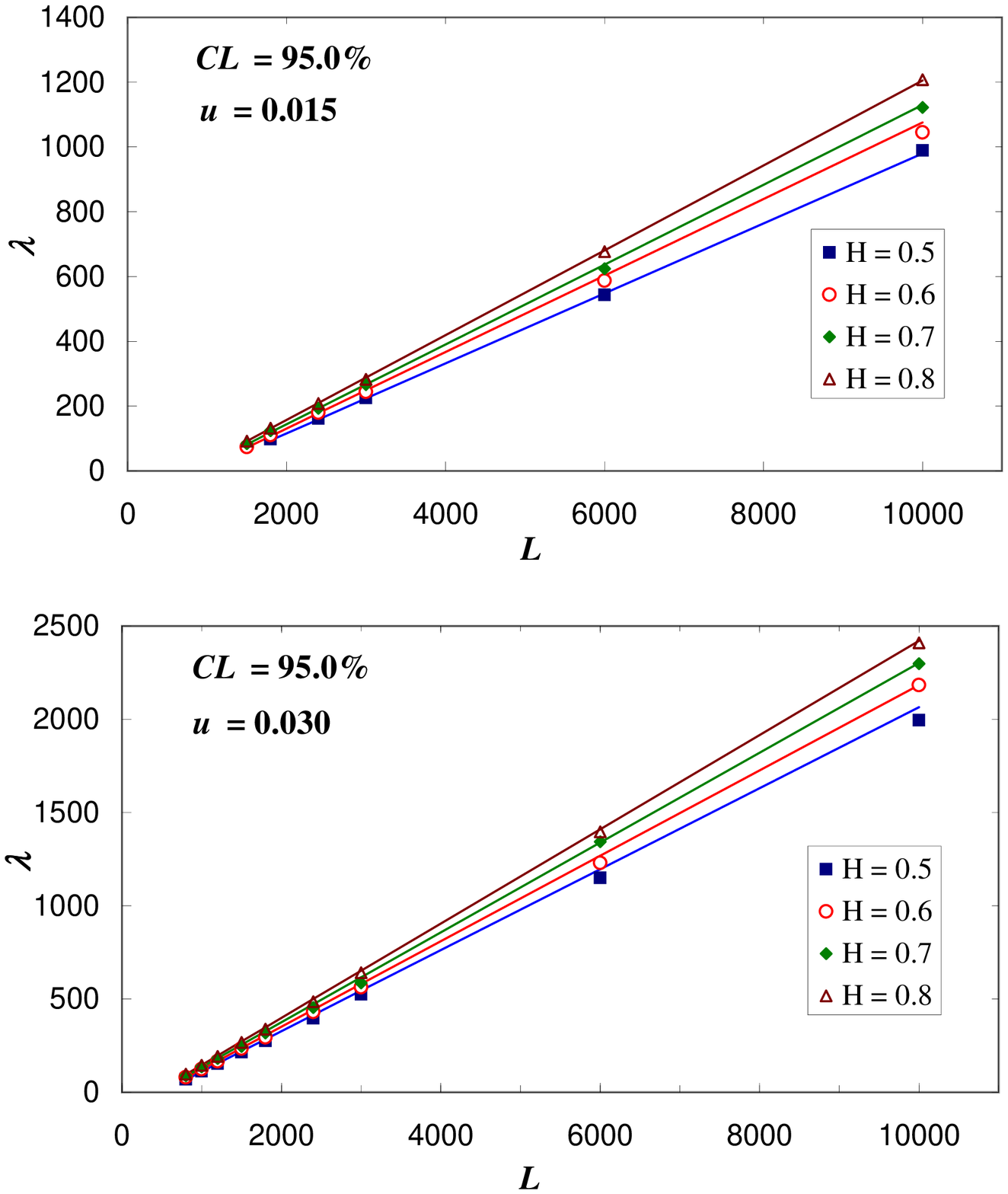,width=15cm,angle=0}}
\end{center}
\caption{Same as in Fig.13 for $CL=95\%$.}
\end{figure}

\begin{figure}
\begin{center}
{\psfig{file=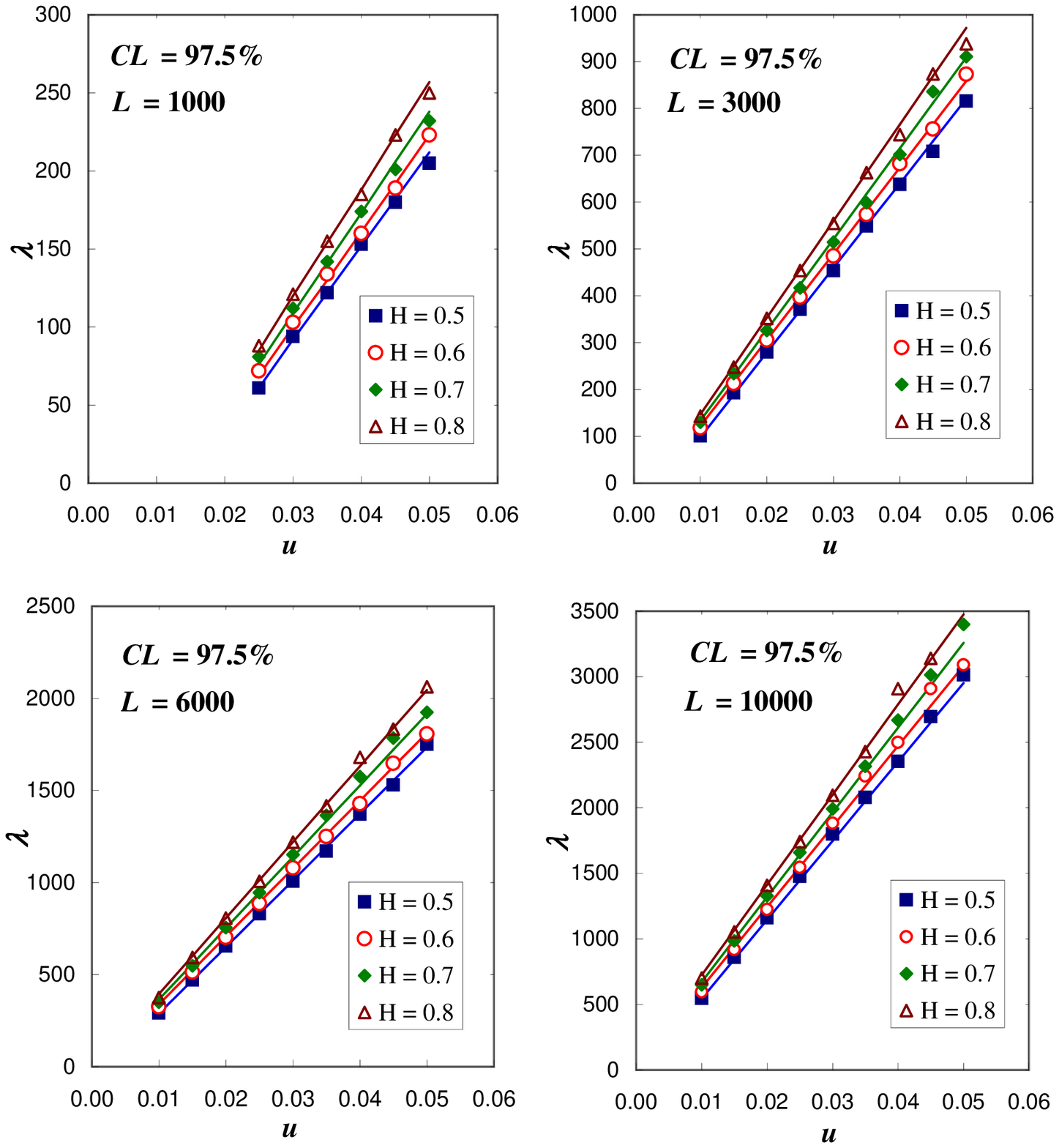,width=15cm,angle=0}}
\end{center}
\caption{Best fit results of Eq.(7) found for simulated series  of autocorrelated data at $CL=97.5\%$ as in Fig.11 and shown as the function of $u$. The cases of several lengths of data are shown. Other $L$ values (not shown due to lack of space) show also linear dependence on $u$. Parameters of the fits are gathered in Table 2.}
\end{figure}

\begin{figure}
\begin{center}
{\psfig{file=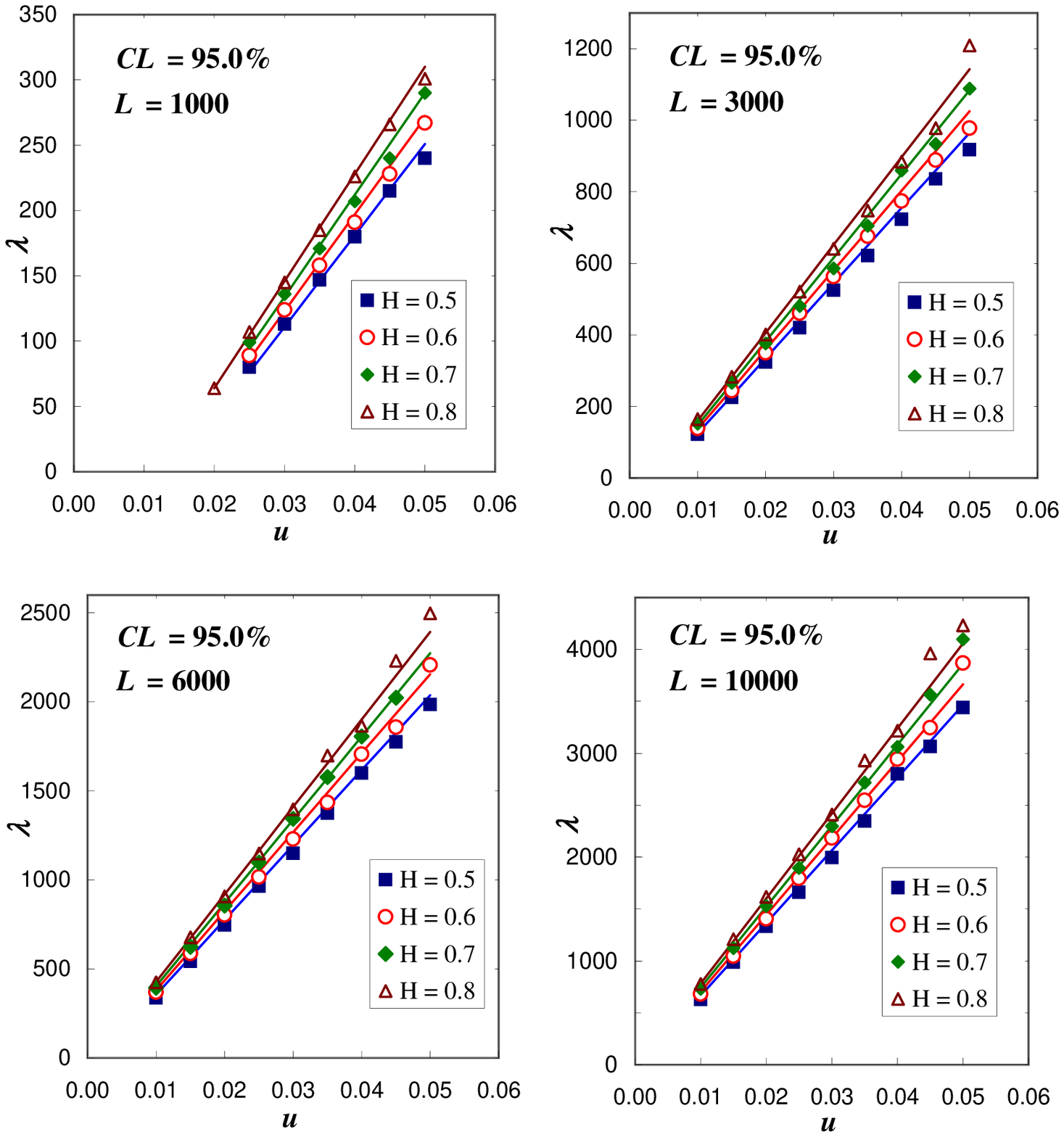,width=15cm,angle=0}}
\end{center}
\caption{Same as in Fig.15, but for $CL=95\%$.}
\end{figure}

\begin{figure}
\begin{center}
{\psfig{file=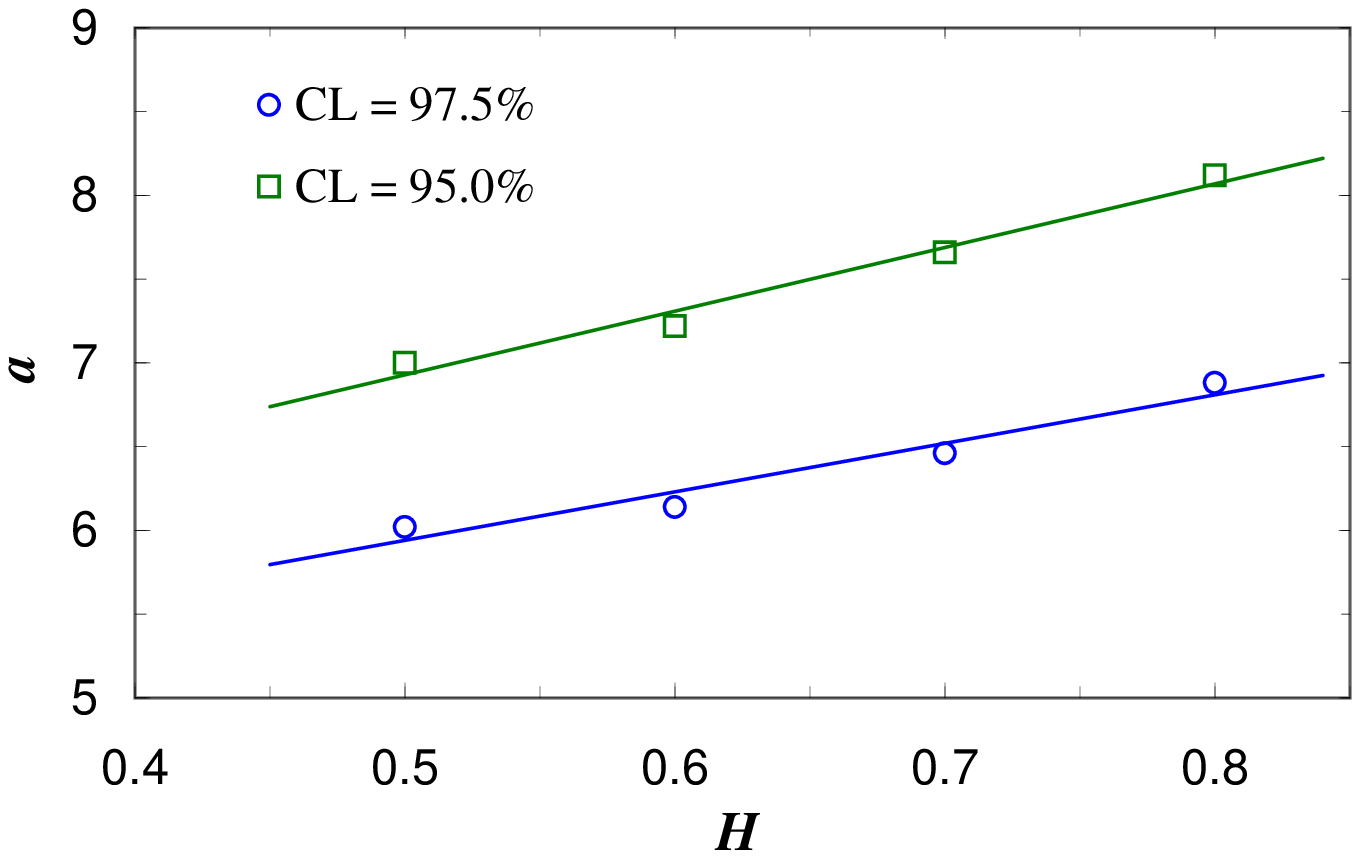,width=14cm,angle=0}}
\end{center}
\caption{Dependence between the fitted $a(H)$ parameter (see Eq.(7)) and the autocorrelation level in data expressed by Hurst exponent $H$.}
\end{figure}

\end{document}